\documentclass[a4paper,fleqn]{cas-dc}

\usepackage[numbers]{natbib}

\def\tsc#1{\csdef{#1}{\textsc{\lowercase{#1}}\xspace}}
\tsc{WGM}
\tsc{QE}
\tsc{EP}
\tsc{PMS}
\tsc{BEC}
\tsc{DE}

\begin{document}
\let\WriteBookmarks\relax
\def\floatpagepagefraction{1}
\def\textpagefraction{.001}
\shorttitle{Three Laws of Technology Rise or Fall}
\shortauthors{Jianfeng Zhan}

\title [mode = title]{Three laws of technology rise or fall}




\author{Jianfeng Zhan}
\ead{zhanjianfeng@ict.ac.cn}
\ead[url]{www.benchcouncil.org/zjf.html}


\address{Research Center for Advanced Computer Systems, Institute of Computing Technology, Chinese Academy of Sciences, China}








\begin{abstract}
Newton's laws of motion perfectly explain or approximate physical phenomena in our everyday life. Are there any laws that explain or approximate technology's rise or fall? After reviewing thirteen information technologies that succeeded, this article concludes three laws of technology and derives five corollaries to explain or approximate the rise or fall of technology. Three laws are the laws of technology inertia, technology change force, and technology action and reaction. Five corollaries are the corollaries of measurement of technology change force, technology breakthrough, technology monopoly, technology openness, and technology business opportunity. I present how to use the laws and the corollaries to analyze an emerging technology---the open-source RISC-V processor. Also, I elaborate on benchmarks' role in applying those laws.
\\
\\
\\
\end{abstract}



\begin{keywords}
Technology laws and corollaries\\ 
Law of technology inertia\\
Law of technology change force\\ 
Law of technology action and reaction\\
Corollary of measurement of technology change force\\
Corollary of technology breakthrough \\
Corollary of technology openness\\ 
Corollary of technology monopoly\\ 
Corollary of technology business opportunity \\
RISC-V
\end{keywords}
\maketitle

\section{Introduction}

Basic research, including both basic scientific research and basic technological research~\cite{national1999funding}, generates scientific and technical knowledge. At the same time, technology is the application of scientific and technical knowledge for practical purposes, especially in the industry~\cite{oxfordlan}. Researchers published many scientific and technical papers or proposed massive new concepts or ideas. Upon scientific and technical knowledge, engineers develop massive technologies, e.g., tools, platforms, products, and services in the information technology (IT) industry. But few technologies matured and succeeded! Why? Are there any laws that can describe or approximate technology rise or fall as Newton's laws of motion perfectly explain or approximate physical phenomena in our everyday life?

\begin{table*}[htbp]
\caption{The summary of laws of technology.}
\centering
\begin{tabular}{|p{7.5cm}|p{8cm}|}
\hline
Law or corollary name & Formula   \\ \hline
\multirow{2}{*}{Law of technology inertia} & $\Delta U_{t}=U_{(t+\Delta t)}-U_{t}$ \\
& $\Delta U_{t}=0\;,\;U\;is\;a\;natural\;number$  \\\hline
Law of technology change force& $\Delta U_{t} \propto F_{t}\;,\;U\;is\;a\;natural\;number$\\ \hline
Law of technology change action and reaction& $F_{Emerging}=-F_{Existing}$\\ \hline
\multirow{3}{*}{Corollary of measurement of technology change force}& $F_{t}=F_{Creating}+F_{Learn}+F_{Ecosystem}$ \\
& or \\
& $F_{t}=F_{Experience}+F_{Cost}+F_{Efficiency}+F_{Other}+F_{Learn}+F_{Ecosystem}$\\ \hline
Corollary of technology breakthrough& $B=F/U\;,\;U\;is\;a\;natural\;number$\\ \hline
\multirow{3}{*}{Corollary of technology openness} & $P=p*U_{i}$, $c=C/U_{i}$\\ 
& $P=p*U_{i}/M$, $c=C*M/U_{i}$\\ 
& $P=p*U_{i}/N$, $c=C*N/U_{i}$  \\\hline
Corollary of technology monopoly& $U=\Sigma \Delta U_{t} \propto F$\\ \hline
Corollary of technology business opportunity& /\\ \hline
\end{tabular}
\label{tab_law}
\end{table*}

The National Research Council of the US ~\cite{national2003innovation,national2012continuing,national2020information} published a series of white papers on the patterns of innovation in IT, evidenced by many case studies. The white papers concluded IT has a long, unpredictable incubation period, and the university and industry have a complex partnership in innovations. Meanwhile, they observed the by-product, resurgence, and confluence phenomena in the IT innovations ~\cite{national2003innovation,national2012continuing,national2020information}. Also, many scholars pondered the nature of technology. Jacques Ellul~\cite{ellul1964technological} and Langdon Winner ~\cite{winner1978autonomous} concluded the philosophical doctrine of technological determinism ~\cite{kranzberg1986technology}. Lynn Townsend White declined to accept this technological omnipotence ~\cite{kranzberg1986technology}. Instead, he claimed that a technical device "merely opens a door, it does not compel one to enter." ~\cite{kranzberg1986technology,white1964medieval}. Melvin Kranzberg ~\cite{kranzberg1986technology} concluded a series of observations deriving from a longtime study, Kranzberg's Six Laws of Technology. These observations, quests, or arguments help understand the development of technology and its interactions with sociocultural change ~\cite{national2003innovation,national2012continuing,national2020information,kranzberg1986technology}. Still, they fail to provide a qualitative or quantitative approach to explain or approximate the rise or fall of technology.

After analyzing thirteen successful IT, I conclude three laws analogous to Newton's laws of motion to explain technology's rise and fall in Section 2. The first law is on the obstacle to new technology: technology inertia. Not only end-users but also industry users stick to the existing technology. The user size will keep constant unless a non-zero net technology change force acts on it. The second law reveals where the power of new technology comes from. The change of user size is proportional to the net technology change force. The corollary of measurement of technology change force is on how to measure the net change force. Only by creating a brand-new technology or improving an existing technology in terms of user experience, costs, efficiency, or other fundamental dimensions by several orders of magnitude  can the new technology generate a positive change force. The third law explains how existing and emerging technologies compete. When the net technology change force is positive, the emerging technology rises and the existing technology falls, or else the existing technology keeps. In Section 3, I derive two corollaries, named the corollaries of technology breakthrough and technology monopoly, to explain how a new technology breakthroughs and gains mono\-poly, respectively. In addition, two corollaries called the corollaries of technology openness, and technology business opportunity explain why an open technology gains edge over a closed one and how a technology achieves business opportunities, respectively. Table~\ref{tab_law} summarizes the three laws and five corollaries together. Table~\ref{tab3} explains the symbols in the formula in Table~\ref{tab_law}.

\begin{table}[htbp]
\caption{The explanations of symbols in Table~\ref{tab_law}.}
\centering
\begin{tabular}{|p{1.5cm}|p{6cm}|}
\hline
Symbol & Explanation   \\ \hline
$\Delta$ & Difference operator \\ \hline
$\propto$ & Proportional operator \\ \hline
$\Sigma$ & Summation operator \\ \hline
$U_{t}$ & User size varying with time \\ \hline
$U_{i}$ & Size of industry users \\ \hline
$F_{t}$ & Net technology change force varying with time \\ \hline
$F_{Emerging}$ & Change force acting on emerging technology \\ \hline
$F_{Existing}$ & Change force acting on existing technology \\ \hline
$F_{Create}$ & Change force resulted from creating a brand-new technology \\ \hline
$F_{Learn}$ & Change force resulted from learning cost \\ \hline
$F_{Ecosystem}$ & Change force resulted from ecosystem deviation \\ \hline
$F_{Experience}$ & Change force resulted from user experience \\ \hline
$F_{Cost}$ & Change force resulted from cost \\ \hline
$F_{Efficiency}$ & Change force resulted from efficiency\\ \hline
$F_{Other}$ & Change force resulted from other fundamental dimensions \\ \hline
$B$ & Technology breakthrough \\ \hline
$P$ & Gross productivity \\ \hline
$C$ & Total cost \\ \hline
$M$ & Number of decoupled supply chains \\ \hline
$N$ & Number of nations \\ \hline
$c$ & Cost of each contributor (industry user) \\ \hline
$p$ & Productivity of each contributor (industry user) \\ \hline
\end{tabular}
\label{tab3}
\end{table}

I want to emphasize that these laws differ from Newton's laws. The former is difficult to verify through quantitative experiments under repeatable or reproducible settings because its settings are highly complex. Some laws stand the shoulders of giants, quoted from Newton's saying. For example, the previous work~\cite{thiel2014zero} presented observations similar to the corollaries of measurement of technology change force and technology monopoly. I point out the difference in each subsequent subsection. My contribution is to propose a simple but powerful theory framework (three laws and five corollaries) to explain the technology's rise or fall.

Section 4 uses the three laws and the four corollaries to predict an emerging technology RISC. Section 5 discusses the role benchmarks played in applying those laws and corollaries. Section 6 presents the related work. Finally, I conclude Section 7.

\section{The three laws of technology}
After analyzing the technology in Table~\ref{tab1}, this section presents the laws of technology inertia, technology change force, and technology action and reaction, respectively.  

\subsection{The law of technology inertia (the first law of technology).}

\begin{figure}
	\centering
		\includegraphics[scale=1]{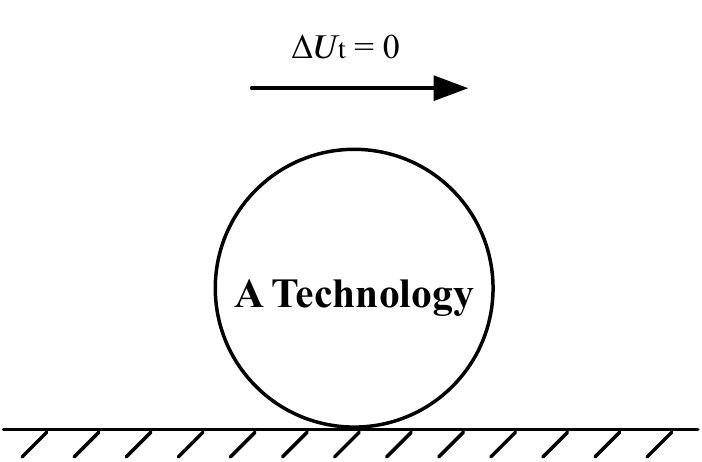}
	\caption{The law of technology inertia.}
	\label{FIG:1}
\end{figure}

Newton's first law states that an object moves with a velocity that is constant in magnitude and direction unless a non-zero net force acts on it~\cite{serway2011college}. Similar to Newton's first law, I presume the existence of the technology inertia - the user size $U_{t}$ of a technology will keep constant unless a non-zero net technology change force $F_{t}$ acts on it.  Out of learning pressure and usage habits, end-users $U_{e}$ stick to existing products, tools, platforms,  and services called consumer inertia. Industrial users $U_{i}$ adhere to existing products, tools, platforms, and services for investment protection, called ecosystem inertia. The consumer inertia and ecosystem inertia together constitute the technology inertia. The user size in the first law is a normalized number referring to both end-users $U_{e}$ and industrial users $U_{i}$. Eq.~\ref{formula-1} quantitatively states this law.  I use a normalized user size that considers the weights of end-users and industrial users. Fig.~\ref{FIG:1} presents the law of technology inertia. The first law is a hypothesis for the general trend. Someone may argue the exception case also holds. I do not doubt it. But the law of technology inertia is a hypothesis at a macroscopic level, not a microscopic level.

\textbf{The law of technology inertia}:  the user size $U_{t}$ will keep constant unless a non-zero net technology change force $F_{t}$ acts on it. 

\begin{equation}
\label{formula-1}
\begin{aligned}
&\Delta U_{t}=U_{(t+\Delta t)}-U_{t}\\
&\Delta U_{t}=0\;,\;U\;is\;a\;natural\;number 
\end{aligned}
\end{equation}

A technology's ecosystem is complex, needing an exhausting analysis. I will elaborate on them in Section 4. 

\subsection{The law of technology change force (the second law of technology)}

\begin{figure}
	\centering
		\includegraphics[scale=1]{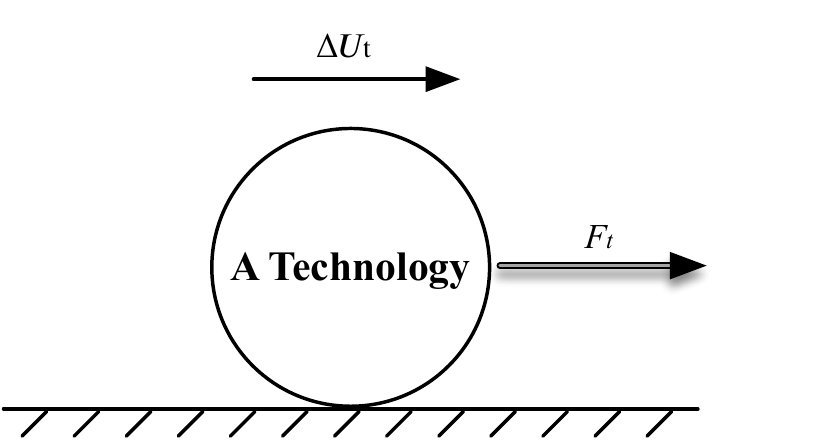}
	\caption{The law of technology change force.}
	\label{FIG:2}
\end{figure}

Newton's second law states that the acceleration of an object is directly proportional to the net force acting on it and inversely proportional to its mass~\cite{serway2011college}. Like Newton's second law, I presume the change of the user size is proportional to the net technology change force, which I call the law of technology change force. Eq.~\ref{formula-2} states this relationship qualitatively. The user size increases when the net change force is positive, while the user size decreases when the net change force is negative. When the net change force is zero, the user size keeps constant according to Eq.~\ref{formula-1}. Please note that in Eq.~\ref{formula-2}, $U$ is a natural number. With a negative change force acting, when $U=0$ decreases to zero, it will trigger a net change force $F_{t}=0$ because a negative user size has no physical meaning. I can not quantify the relationship between $\Delta U$ and $F_{t}$ accurately as it is much more challenging. 
Fig.~\ref{FIG:2} presents the law of technology change force.

\textbf{The law of technology change force}: The change of user size $\Delta U_{t}$  is proportional to the net technology change force $F_{t}$.

\begin{equation}
\label{formula-2}
\begin{aligned}
&\Delta U_{t} \propto F_{t}\;,\;U\;is\;a\;natural\;number
\end{aligned}
\end{equation}

Creating a brand-new technology will generate a positive change force $F_{create}$. However, from proposing a new concept to falling into the ground, it has a long, unpredictable incubation period separating into multiple phases of innovation: research exploration, initial commercial deployment, and eventual business breakthrough~\cite{national2003innovation,national2020information}. My law of technology change force gives another explanation. Though creating a brand-new technology will generate a positive cha\-nge force, the learning cost (consumer inertia) of end-users $F_{Learn}$ and ecosystem inertia of industry users  $F_{Ecosystem}$ will generate negative change forces. As these different components counter each other, the net change force will vary and finally decide its rise or fall. The steam engine, plane, and computer industries all witnessed this process. Table~\ref{tab1} presents IT examples like WWW, Internet, Google, Facebook. 

Melvin Kranzberg~\cite{kranzberg1986technology} explains the nature of creating a brand-new technology in Kranzberg's Second Law: Invention is the mother of necessity. He took the automobile as an example of how a successful technology requires auxiliary technologies to make it fully effective~\cite{kranzberg1986technology}: the automobile brought whole new industries by their need for rubber tires, petroleum products, and new tools and materials;  Large-scale commercial automobile deployment demanded roads, highways, garages, parking lots, traffic signals, and parking meters~\cite{kranzberg1986technology}.

For an existing technology, only improving its user experience, cost, efficiency by several orders of magnitude can generate a positive change force $F_{Improve}$ that can break thro\-ugh the technology inertia, or else it will generate a negative change force. Thiel et al. proposed a rule of thumb in~\cite{thiel2014zero}: Proprietary technology must be at least ten times better than its closest substitute in some important dimension to lead to real monopolistic advantage. Otherwise, it will probably be perceived as a marginal improvement. Table~\ref{tab1} presents IT examples like deep learning, RAID, etc. 

In improving an existing technology, a new or different ecosystem will generate a negative change force $F_{Ecosystem}$, which is the side effect of ecosystem inertia. Meanwhile, different use which results in a learning cost will generate a negative change force $F_{Learn}$. When two rivals can substitute without causing any overhead for end-users and industrial users, the net change force is zero, and the earlier technology gets a head start. I conclude the above discussion as a corollary, named the law of measurement of technology change force. Eq.~\ref{formula-3} states the net change force $F_{t}$ is the sum of $F_{Create}$, $F_{Learn}$ and $F_{Ecosystem}$ or the sum of $F_{Improve}$, $F_{Learn}$ and $F_{Ecosystem}$.

\textbf{Corollary of Measurement of Technology Change For\-ce}: Only by creating a brand-new technology ($F_{Create}$)  or improving existing technology ($F_{Improve}$) in terms of user experience ($F_{Experience}$), cost ($F_{Cost}$), efficien\-cy ($F_{Efficiency}$), or other fundamental dimensions ($F_{Other}$) by several orders of magnitude can it generate a positive net change force, or else it will generate a negative change force. A new ecosystem or the deviation between emerging and existing technology ecosystems will generate a negative change force  $F_{Ecosystem}$. Different use, which results in an end-user learning cost, will generate a negative change force $F_{Learn}$. The net technology change force is zero when two rivals can substitute without causing any overhead for end-users and industrial users.  

\begin{equation}
\begin{aligned}
\label{formula-3}
&F_{t}=F_{Create}+F_{Learn}+F_{Ecosystem}    \\
&or \\
&F_{t}=F_{Experience}+F_{Cost}+F_{Efficiency}+F_{Other}+F_{Learn}
\\&+F_{Ecosystem} 
\end{aligned}
\end{equation}

\subsection{The law of technology action and reaction (the third law of technology)}~\label{subsection_third_law}

\begin{figure*}
	\centering
		\includegraphics[scale=1]{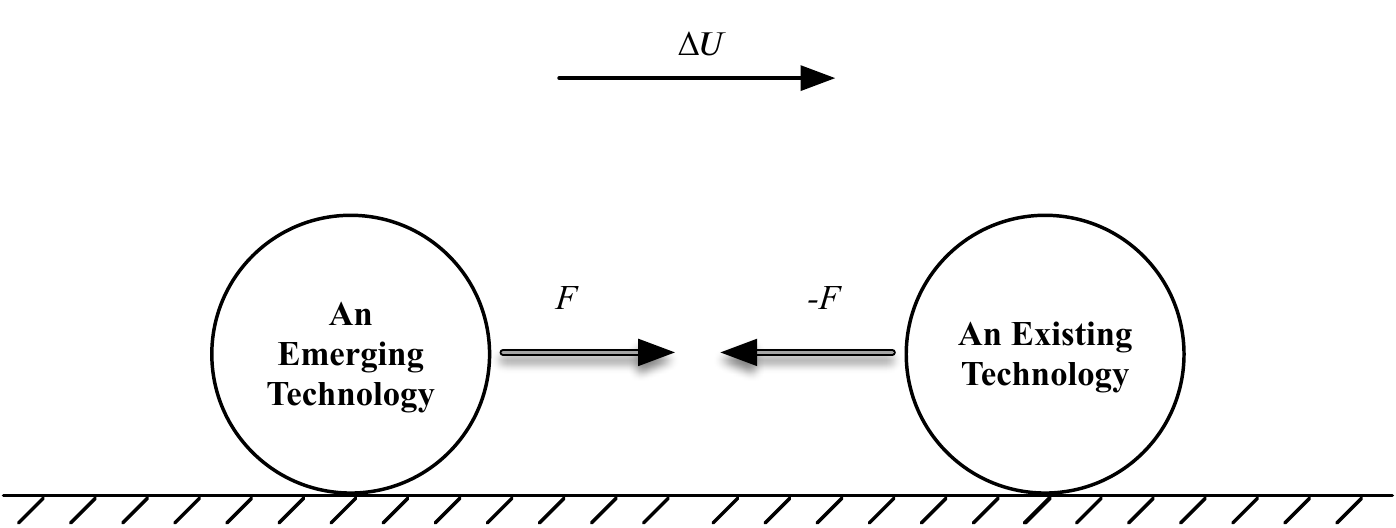}
	\caption{The law of technology action and reaction. In this figure, the technology change force is positive. }
	\label{FIG:3}
\end{figure*}

Newton's third law states that if object one and object two interact, the force $F_{12}$ exerted by object one on object
two is equal in magnitude but opposite in direction to the force $F_{21}$ exerted by
object 2 on object 1~\cite{serway2011college}. Similar to Newton's third law, I state the third law of technology: technology change forces acting on the emerging technology ($F_{E}$
$_{-merging}$)  and reacting on the existing technology ($F_{Existing}$)  equal in magnitude but opposite in direction. Fig.~\ref{FIG:3} prese\-nts the law of technology action and reaction. A change force acts on the emerging technology while a change force in the same magnitude but different directions reacts to the existing technology. When the change force acting on the emerging technology is positive, the emerging technology rises (user size increases), and the existing technology falls (user size declines). Eq.~\ref{formula-4} states this relationship quantitatively.  

\textbf{The law of technology action and reaction}:
Under the normal conditions, technology change forces acting on the emerging technology ($F_{Emerging}$)  and reacting on the existing technology ($F_{Existing}$)  are equal in magnitude but opposite in direction. 

\begin{equation}
\label{formula-4}
\begin{split}
&F_{Emerging}=-F_{Existing}
\end{split}
\end{equation}

Furthermore, I clarify the normal conditions where the third law of technology holds. First, it should exclude decoupling of the supply chain, which the pandemic accelerates. The decoupling of the supply chain is why the same technology is replicated, or superior and inferior technology coexist in different nations. Second, it should exclude the case where the government has total control over the production and pricing of goods and services. In the latter case, the market will weaken the technology change forces. Third, in a normal condition, it must warrant the protection of intellectual properties and the punishment of business fraud. Superior technology can be mirrored or stolen without paying prices without protecting intellectual properties. Meanwhile, tolerating business fraud weakens the signals emitted by technology change forces.  

I take two computer examples in the shift from server to desktop to mobile  to explain it. The creation of standardized, vendor-independent operating systems, such as UNIX and its open-source clone, Linux, shields the new architecture's change, making the ecosystem inertia against new architecture disappear. And hence it lowered the cost and risk of bringing out a new architecture ~\cite{hennessy2011computer}.  According to Corollary One, $F_{Ecosystem}=0$.

The RISC-based computers raised the performance bar, and its efficiency generate a positive change force ($F_{Efficien}$
$_{-cy}>0$) that  broke through the technology inertia of the prior architectures like Digital Equipment VAX, forcing the latter to disappear~\cite{hennessy2011computer} (RISC vs. VAX). The late 1990s witnessed the soaring transistor counts, which provided a survival chance for Intel X86 architecture. As the hardware overhead becomes negligible, the latter translates 80x86 instructions into RISC-like ones internally to keep up with the RISC ecology---leveraging many of the innovations first pioneered in the RI\-SC designs and pacing with the RISC performance ($F_{Ecosystem}=0$ and $F_{Efficiency}=0$). In this context, the RISC architecture generates an almost zero net change force against the X86 architecture (RISC vs. X86) ($F=0$). X86 survive. After the emergence of mobile applications, the constraints of power and silicon area make the x86-translation overhead unacceptable and lead to the dominance of a RISC architecture, ARM~\cite{hennessy2011computer}. ARM's reducing cost ($F_{Cost}>0$) and improving efficiency ($F_{Efficiency}>0$) while compatible with the ecosystem ($F_{Ecosystem}=0$)  generate a positive chan\-ge force that can break through the technology inertia of x86 (ARM vs. x86).

Microsoft and other companies predicted the shift from desktop to mobile. The competition between Windows Mobile and Android is governed by the law of technology change force. 
The Android OS bases Linux, which worldwide world-class engineers contribute, offering the open-source operating system to third-party mobile manufacturers for free~\cite{hist2021android}. In contrast, Windows Mobile chose the opposite direction: closed-source Windows as the basis and non-free use (Android is free, while Windows Mobile costs manufacturers \$15 to \$25 a phone in 2009 when Windows Mobile dominated Android ~\cite{hansell2009big}). Meanwhile, Android is designed for mobile:  Android's software is intended for modern screens you tap with a finger, while Windows Mobile was built for use with a stylus ~\cite{hansell2009big}. Both cost ($F_{Cost}>0$) and user experience ($F_{Experience}>0$) edges of Android shook the technology inertia, while Windows Mobile faded (Android vs. Windows Mobile).

Social networking examples also confirm this law. Social networking applications like Facebook, Twitter, WeChat fell into the ground, bringing devastating advantages and benefits. Many companies want to copy the success. For example, several E-commerce service providers and mobile phone giants with massive user bases, strong financial support, and world-class engineers are devoted to building social-networking applications. Predictably, they all failed.    
Out of learning pressure and usage habits, users stick to Facebook, Twitter, or WeChat. Moreover, Facebook, Twitter, or WeChat open their services and API to third parties to build the ecosystem. The third-parties service providers also are accustomed to adhering to Facebook, Twitter, or WeChat for investment protection.
Why are there similar social networking applications in the US and China? I discuss this case above.

\begin{table*}[htbp]
\caption{The analysis of thirteen successful IT.}
\centering
\resizebox{\textwidth}{!}{
\begin{tabular}{|c|c|c|c|c|c|c|c|c|c|c|}
\hline
Technology & Rivals &$F_{Learn}$ & $F_{Ecosystem}$ & $F_{Create}$ & $F_{Cost}$ & $F_{Efficiency}$ & $F_{Experience}$ & $F_{Other}$ & Closed/Open & Supply chain  \\ \hline
Deep learning  & Shallow neutral networks &0& <0 & /  & <0 & / & / & >0 (accuracy) & Open & Coupling  \\ \hline
WWW            & No &<0 & <0 & >0 & / & / & / & / & Open & Coupling  \\ \hline
Google  & No & <0  &<0& >0 & / & / & / & / & Closed & Decoupling \\ \hline
Facebook  & No &<0 &<0 & >0 & / & / & / & / & Closed & Decoupling  \\ \hline
Internet  & No &<0 & <0 & >0 & / & / & / & / & Open & Coupling  \\ \hline
RAID  & Single large expensive disk &0 & <0 & / & >0 & >0 & / & / & Open & Coupling  \\ \hline
Android  & Windows Mobile, Symbian, iOS, Linux &<0 & 0(with respect to Linux) &/ & >0 & / & >0 & >0 (Google ecosystem)& Open & Coupling  \\ \hline
iOS  & Windows Mobile, Symbian &<0 & <0 & / & <0 & / & >0 & >0 (AppStore) & Closed & Coupling  \\ \hline
Windows  & DOS &<0& 0 & / & / & / & >0 & / & Closed & Coupling  \\ \hline
Linux  & UNIX &0 & 0 & / & >0 & / & / & / & Open & Coupling  \\ \hline
UNIX  & Multics & <0 & <0 &/ &/ & >0& / & >0 (standard) & Closed & Coupling  \\ \hline
ARM  & X86, RISC & 0 &0(with respect to RISC) & / & / & >0 & / & >0 (energy efficiency) & Closed  & Decoupling  \\ \hline
RISC  & CISC &0& 0 & /& / & >0 & / & / & Closed & Decoupling \\ \hline
\end{tabular}}
\label{tab1}
\end{table*}

\section{Four corollaries of technology}
I discuss several corollaries as follows. 

\subsection{Corollary of technology breakthrough}

\textbf{Corollary of technology breakthrough}: The technology breakthrough $B$ is defined as the net change force $F$ divided by the target or real user size $U$.  There are three ways to raise technology breakthrough: increase the net change force, satisfy a smaller target users' requirement through focusing on a more minimal set of functions, ensure the emerging technology's ecosystem is compatible with the existing one. 

\begin{equation}
\label{formula-5}
\begin{aligned}
&B=F/U\;,\;U\;is\;a\;natural\;number
\end{aligned}
\end{equation}

\begin{figure}
	\centering
		\includegraphics[scale=1]{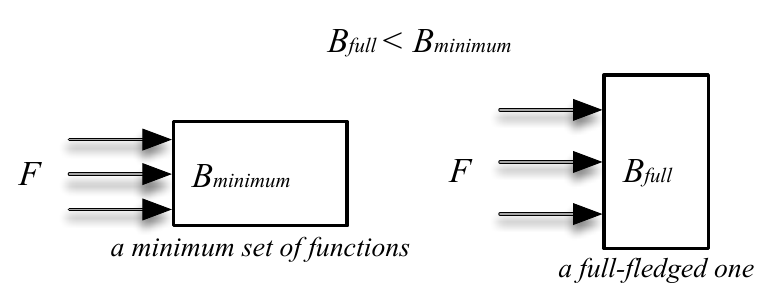}
	\caption{The corollary of technology breakthrough.}
	\label{FIG:4}
\end{figure}

I use a physics concept, Pressure $P$, to explain this law. If $F$ is the magnitude of a force exerted perpendicular to a given surface of
area $A$, then the average Pressure $P$ is the force divided by the area~\cite{serway2011college}.
When a new technology emerges, it needs to overcome the technology inertia and contact end-users $U_{e}$ and industry users $U_{i}$. Here, I use the user size $U$ to measure the given surface of the contact area. So, according to the pressure definition, the breakthrough $B$ is defined in Eq.~\ref{formula-5}. The increase of the net change forces will raise the breakthrough.  Ensuring the emerging technology compatible with the ecosystem of the existing technology will ensure $F_{Ecosystem}=0$ according to the corollary of measurement of technology change force. 

Focusing on a minimum set of functions compared to a full-fledged one has several benefits. First, a smaller contact area $A$ (smaller target users $U$) with the same net change force will exert a larger pressure (breakthrough $B$). Second, it is much easier to improve user experience, cost, or efficiency (raise the net change force). Last but not least, it will control the cost, which is also paramount for innovation. Of course, a minimum set of functions is the lower bound of granularity, or else too simple technology cannot satisfy the users' basic requirements.

I still take the mobile operating system as an example. When iOS and Android emerged, Symbian and Windows Mobile dominated the mobile operating system market. In January 2007, Steve Jobs announced the original iPhone, a touch-based smartphone without a physical keyboard, controlled with a finger without needing a stylus~\cite{ori2007iphone}. The first version of iOS (no official name) was quite limited, focusing on the fresh user experience of introducing a complete touch screen. Contrasting, most users are accustomed to typing on a hardware keyboard at that time. The first iOS version lacked many features those alternative systems already had~\cite{simon2017iphone}. 

While the first version of the iPhone featured a closed ecosystem with the strength of clean and simple design, the first Android phone built the edges of both cost and user experience, brought a free and open-source ecosystem, personality, and the ability to customize user experience~\cite{jess2018phone}. The first version of Android features several unique characteristics~\cite{jess2018phone} (1) choose open-source Linux as a basis, and it is free; (2) allow both the third parties and users to customize the systems, and even additional applications and features; (3) launch Android Market (Google Play) to build application ecosystem; (4) integrate the Google’s already robust ecosystem, constantly expanding software. Android and iOS built full-fledged functions later. Those two typical examples witnessed the law of technology breakthrough. 

\subsection{Corollary of technology monopoly}

\textbf{Corollary of technology monopoly}: The grea\-ter the net technology change force $F$, the higher the technology mono\-poly.

According to the second law, the change of user size $\Delta U$ is proportional to the net technology change $F$.  For the same time interval $T$, the greater the net technology change force, the higher user size $U$=$U_{t}+\Delta U$. Eq.~\ref{formula-6} states this relationship. Naturally, much superior technology gains a higher technology monopoly. Please note that in Eq.~\ref{formula-6}, $F$ is a normalized value as it may vary in different intervals.  In~\cite{thiel2014zero}, monopoly means the kind of company that is so good at what it does that no other firm can offer a close substitute.  Currently, Android and iOS gain the technology monopoly am\-ong the mobile operating systems. 

\begin{equation}
\label{formula-6}
\begin{aligned}
&U=\Sigma \Delta U_{t} \propto F
\end{aligned}
\end{equation}

\subsection{Corollary of technology openness}

\textbf{Corollary of technology openness}: Cont\-rasted with a closed ecosystem controlled by one entity, allowing the division of labor among contributors who share an open technology ecosystem improves the gross productivity $P$  and lowers the cost $c$ amortized on each contributor ($U_{i}$).  Decoupling the supply chain ($M$) and export control of technology between nations($N$) will increase the cost and lower productivity. Standards will decrease the cost of collaboration in industry contributors.

\begin{equation}
\label{formula-7}
\begin{aligned}
&P=p*U_{i}\\
&c=C/U_{i}
\end{aligned}
\end{equation}

\begin{equation}
\label{formula-8}
\begin{aligned}
&P=p*U_{i}/M\\
&c=C*M/U_{i}
\end{aligned}
\end{equation}

\begin{equation}
\label{formula-9}
\begin{aligned}
&P=p*U_{i}/N\\
&c=C*N/U_{i}
\end{aligned}
\end{equation}

There are $N$ nations, $M$ decoupled supply chains, $U_{i}$ technology contributors. $N, M, U_{i}$ are natural numbers. For a specific technology, the gross productivity is $P$, and the total cost is $C$. I presume there is the same size of the technology contributors in each nation or decoupled supply chain. The productivity of each contributor is the same, $p\;jobs\;per$
$day$. As the goal is to perform qualitative analysis, this presumption is reasonable. Eq.~\ref{formula-7} states the total productivity $P$  and the cost of each contributor ( in total, $U_{i}$ contributors). Eq.~\ref{formula-8} states the gross productivity $P$  and the cost of each contributor within each decoupled supply chain (in total,  $M$ decoupled supply chains). Eq.~\ref{formula-9} states the total productivity $P$  and the cost of each contributor in the nation (in total, N nations) subject to export control.

The explanation is also intuitive. Allowing the division of labor among contributors, each contributor performs the job in parallel, hence lowering the cost on average. Meanwhile, the gross productivity will improve despite the collaboration overhead. Eq.~\ref{formula-7} shows that the gross productivity $P$ is proportional to the number of contributors $U_{i}$, and the cost of each contributor $c$ is inversely proportional to the number of contributors $U_{i}$. That is why many industries open their services and API to third parties to build the ecosystem.

Decoupling the supply chain ($M$)  will increase the cost and lower productivity because several similar technologies have to been replicated. Eq.~\ref{formula-8} shows that the gross productivity $P$ of each decoupled supply chain is proportional to the number of contributors $U_{i}$ and inversely proportional to the the number of decoupled supply chain $M$;  the cost of each contributor $c$ is inversely proportional to the the number of contributors $U_{i}$ and proportional to the number of decoupled supply chain $M$.  
 
Putting a nation subject to export control ($N$)  will increase the cost and lower productivity of that nation because the similar technology have to been developed by one nation independently. Eq.~\ref{formula-9} shows that the gross productivity $P$ of the nation under export control is proportional to the number of contributors $U_{i}$ and inversely proportional to the the number of nations $N$;  the cost of each contributor $c$ within the nation under export control is inversely proportional to the the number of contributors $U_{i}$ and proportional to the the number of nations $N$.

Standards will decrease the cost of collaboration in industry contributors. The standardization of interfaces, modularity, and interoperability accelerates the technology impact, enables the absorption of innovations into diverse sectors, and propels industry investment by creating network effects~\cite{national2020information}. Metcalfe’s law~\cite{metcalf1993metcalf,metcalfe2013metcalfe} can be used to quantify the network effects. The law states that the value $V$ of a network is proportional to the square of the size $n$ of the network~\cite{metcalfe2013metcalfe}. A classic example of the power of well-designed interfaces is the TCP/IP suite of Internet protocols~\cite{national2020information}.

\subsection{Corollary of technology business opportunity}

\textbf{Corollary of technology business opportunity}: There are four technology business opportunities --- (1) create a brand-new technology; (2) achieve the superior edge of cost, efficiency, user experience, or other fundamental dimensions; (3) seek a superior position in the division labor sharing a technology ecosystem; (4) opportunities arising from supply chain decoupling or export control.

This corollary summarizes the points I discussed above together. I will not repeat all of them. I only elaborate on the third opportunity. The third law of Kranzberg's Six Laws of Technology~\cite{kranzberg1986technology} reads as follows: technology comes in packages, big and small. It has been extended even further by Thomas P. Hughes. Hughes more precisely and accurately uses systems instead of packages, which he defined systems as coherent structures composed of interacting, interconnected components~\cite{Thomas_book}. In seeking a superior position in the division of labor sharing a technology ecosystem, the goal is to develop a superior component that gains the edge of cost, efficiency, user experience, or other fundamental dimensions against other rival components.

\section{Analyzing RISC-V}

\begin{table*}[htbp]
\centering
\caption{The analysis of the RISC-V processor. As Figure~\ref{processor_ecosystem} shows, RISC-V ISA is only a part of the RISC-V processor. The open-source license of RISC-V ISA only ensures it is partially open and coupling. A completely open and coupling RISC-V processor needs the open-source implementation of SoC, shown in Figure~\ref{processor_ecosystem}, without violating the intellectual proprieties.}
\resizebox{\textwidth}{!}{
\begin{tabular}{|c|c|c|c|c|c|c|c|c|c|c|c|}
\hline
Technology & Domain & Rivals &$F_{Learn}$ & $F_{Ecosystem}$ & $F_{Create}$ & $F_{Cost}$ & $F_{Efficiency}$ & $F_{Experience}$ & $F_{Other}$ & Closed/Open & Supply Chain  \\ \hline
\multirow{4}{8em}{RISC-V processor} & Desktop & X86 &0 & Uncertain & / & >0 & Uncertain &Uncertain & Uncertain & Partially open & Partially coupling  \\
\cline{2-12}
& Server & X86 &0 & Uncertain & / & >0 & Uncertain &Uncertain & Uncertain & Partially open & Partially coupling  \\
\cline{2-12}
& Smart Phone  & ARM &0 & Uncertain & / & >0 & Uncertain &Uncertain & Uncertain & Partially open & Partially coupling  \\
\cline{2-12}
& IoT & No & 0 & Uncertain &/ & >0 & Opportunity & Opportunity & Opportunity & Partially open & Partially coupling  \\
\hline
\end{tabular}}
\label{tab2}
\end{table*}

\begin{figure}
	\centering
		\includegraphics[scale=.4]{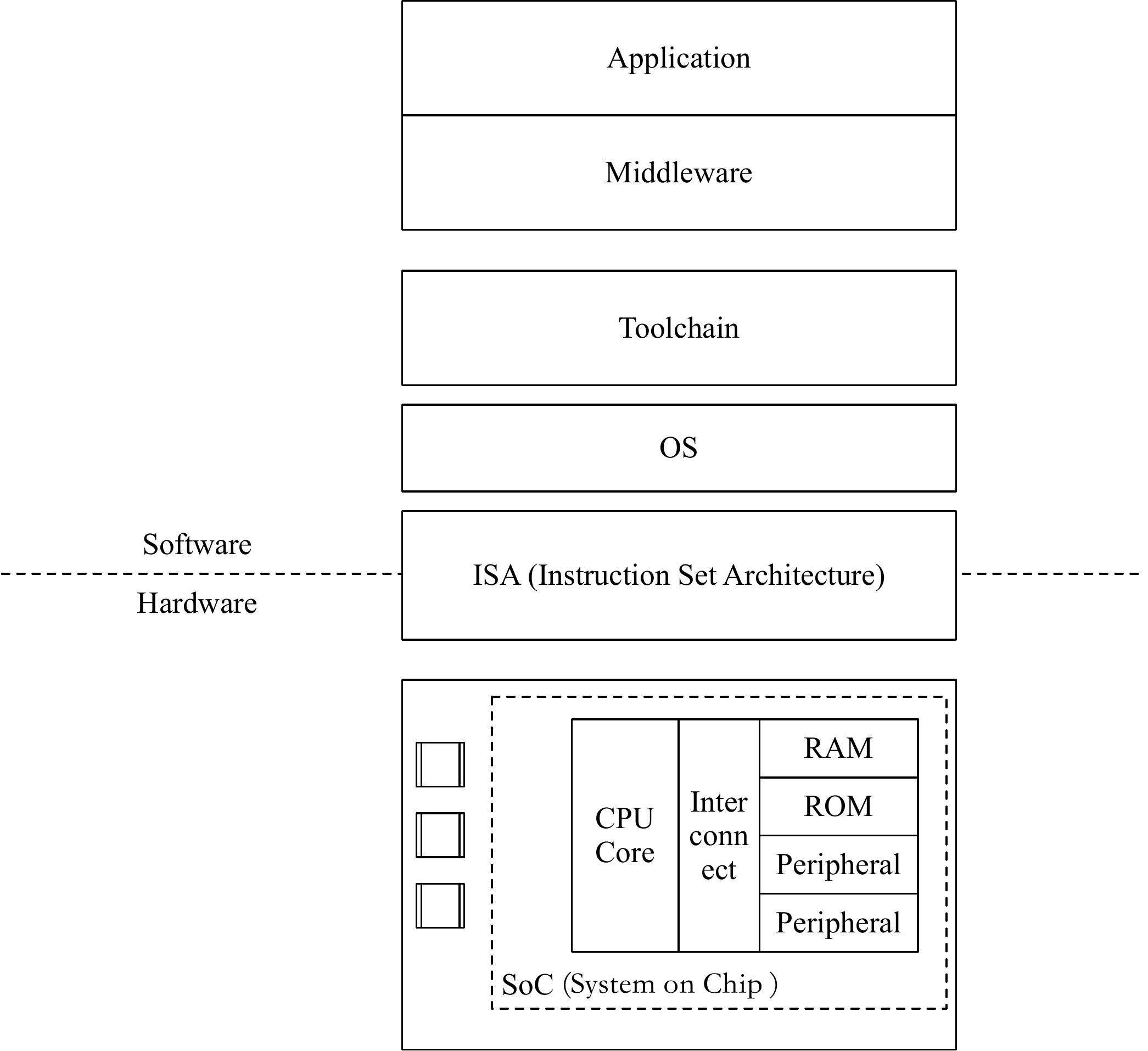}
	\caption{The ecosystems of X86, ARM or RISC-V processors.}
	\label{processor_ecosystem}
\end{figure}

RISC-V~\cite{waterman2014risc} is an open standard instruction set architecture (ISA) that is based on established reduced instruction set computer (RISC) principles, which is provided under open source licenses, unlike most other ISA designs. In this section, I perform a simple analysis of the edge of RISC-V processors against other rivals in different domains (see Table~\ref{tab2}).

Figure~\ref{processor_ecosystem} presents the ecosystem of an X86, ARM, or RISC-V ecosystem. An entire ecosystem of X86, ARM, or RISC-V consists of SoC (a system on a chip), ISA (Instruction Set Architecture), OS (operating system), toolchain, middleware, and applications. SoC is an IC that integrates multiple system components onto a single chip. It often includes CPU cores, memory, input/output ports, and secondary storage, alongside other components such as a graphics processing unit (GPU)~\cite{flynn2011computer}. The Instruction Set Architecture (ISA) defines the interface between software and hardware. The software applications' code is converted to machine instructions, executed at the hardware level. The OS is the primary software managing the hardware resource of a computer. The toolchain is a set of programming tools to develop software, consisting of a compiler, a linker, libraries, and a debugger. Middleware provides services to applications beyond the OS, such as the database system. The application offers the service for the user. 

In the desktop, server, and smartphone domains, rivals like x86 and ARM are mature. Suppose the RISC-V ecology can provide full-fledged functions similar to X86 or ARM, according to Law of Measurement of Technology Change Force,  $F_{Ecosystem}=0$. Unfortunately, it is not a trivial job. According to the law of measurement of technology change force, there are four other components of technology change force: $F_{Cost}$, $F_{efficiency}$, $F_{Experience}$, $F_{Other}$. $F_{Cost}$ is larger than zero. But the other components are uncertain, and they may be negative. In that case, the net change force may be negative. Even the net change force is positive, in coping with this daunting challenge, the law of technology breakthrough reminds us that it is helpful to propose a minimum set of functions to generate a big breakthrough $B$, overcoming the inertia barrier.

The IoT domain is emerging, and the current solutions are fragmented.   RISC-V processor gains a better position as there are no cemented rivals. It is much possible that RISC-V gains edge of efficiency, experience and other fundamental dimensions.  The RISC-V processor has a promising future as it is open and free.

The other advantage of RISC-V is they are partially open and coupling while the other rivals are closed or decoupling. According to the law of technology openness, the division of labor among contributors significantly improves gross productivity and lowers the cost amortized on each contributor. Suppose RISC-V can have a positive net change force (see the above analysis), attracting more contributors ($U_{i}$) joining. The gross productivity $P$ and the cost of each contributor $c$ will increase and decrease with $U_{i}$, respectively. The other pro of RISC-V is that it will face supply chain decoupling and export control to a lesser extent (the SoC may be subject to decoupling supply chain and export control). As the law of technology openness put, decoupling supply chain and export control will increase the cost and lower productivity.

In a word, according to the corollary of technology business opportunities: there are two primary opportunities for RISC-V processors. (1) achieve the superior edges of cost, efficiency, user experience, or other fundamental dimensions; (2) opportunities arising from supply chain decoupling or export control.

\section{The role of benchmark played}
In IT, there are two categories of benchmarks that can play a role in evaluating technology rise or fall ~\cite{ZHAN2021100012}: the first category of the benchmark---a measurement standard and the second one---the representative workloads that run on the systems under measurement. In the other technology domain, a similar methodology like the fifth category of benchmark---benchmarking can be developed ~\cite{ZHAN2021100012}. In this context, benchmarking is the continuous process of searching the industry best practices that lead to superior performance and measuring products, services, and processes ag\-ainst them~\cite{ZHAN2021100012,zairi1996origins}.   

When applying the second technology law, it is best to propose a benchmark to compare against each technology's cost, efficiency, user experiences, and other fundamental dimensions, which is objective.

\section{Related work}

The National Research Council of the US~\cite{national2003innovation,national2012continuing,national2020information} published a series of white papers on the innovation in IT. The first observation is that IT has a long, unpredictable incubation period separating into multiple phases of innovation: research exploration, initial commercial deployment, and eventual business breakthrough~\cite{national2003innovation,national2020information}. A striking example is the field of number theory: a branch of pure mathematics without applications for hundreds of years now became the basis for the public-key cryptography that underlies information security~\cite{national2003innovation}.

The second observation is that the university and industry have a complex partnership in innovations ~\cite{national2003innovation,national2012continuing,national2020information}. One pattern is that the initial ideas came from the industry, not commercialized until the university launched the research projects. For example, IBM pioneered both the concept of reduced-instruction-set computing (RISC) processors and relational databases (its System R project). The former was not commercialized until the University of California at Berkeley and Stanford University performed its Very Large-Scale Integrated Circuit (VLSI) program of the late 1970s and early 1980s~\cite{national2003innovation,national2012continuing,national2020information}. The latter was not commercialized until the University of California at Berkeley brought this technology to where several start-up companies commercialized it~\cite{national2003innovation,national2012continuing,national2020information}. The other pattern is that the initial ideas came from the university community, followed by industry research. The time-sharing system, which aims at making it possible to share expensive computing resources among multiple simultaneous interactive users, follows this pattern ~\cite{national2003innovation,national2012continuing,national2020information}.

The third is the by-product phenomena, where collateral results, often unanticipated, are the by-product of the anticipated ones but as important as the anticipated results of research  ~\cite{national2003innovation,national2020information}. For example, electronic mail and instant messaging were by-products of the time-sharing systems.

The fourth is about the resurgence phenomena. Researc\-hers' interests can fall off in areas where progress has slowed, followed by a resurgence when new ideas or enablers emerge ~\cite{national2020information}. It is difficult to predict a priori which research will pay off rapidly and take time, which typical resurgence examples like Machine Learning, Formal methods, Virtual machines (VMs), and Virtual reality (VR) witnessed~\cite{national2020information}.  

The fifth is about the confluence phenomena where multiple threads of IT-enabled innovation coming together have a transformative impact on a significant industry sector thr\-ough converging contributions from various areas of IT innovation~\cite{national2020information}. The impact of the Internet is the underlying canonical component that contributes to many confluence succes\-ses~\cite{national2020information}. Confluence relies on several factors~\cite{national2020information}: the ability to combine deep expertise about the domain in which IT is being applied with deep knowledge in IT, design and production knowledge that combines knowledge of the application and IT, and the development of new business models that take advantage of the capabilities afforded by IT.

Many scholars pondered the nature of technology. Jacq\-ues Ellul~\cite{ellul1964technological} and Langdon Winner~\cite{winner1978autonomous} conclude the philosophical doctrine of technological determinism~\cite{kranzberg1986technology}: technology is pursued for its own sake and without regard to human need. Melvin Kranzberg~\cite{kranzberg1986technology} interpreted this proposition as "technology has become autonomous and has outrun human control; in a startling reversal, the machines have become the masters of man."
However, some scholars declined to accept this technological omnipotence~\cite{kranzberg1986technology}. Lynn Townsend White claimed that a technical device "merely opens a door, it does not compel one to enter"~\cite{kranzberg1986technology,white1964medieval} . Melvin Kranzberg~\cite{kranzberg1986technology} concluded a series of observations deriving from a longtime study, Kranzberg's Six Laws of Technology. Kranzberg's First Law reads as follows: technology is neither good nor bad; nor is it neutral, which can be interpreted in that the same technology can have quite different results when introduced into different contexts or under different circumstances~\cite{kranzberg1986technology}.

A lots of previous work studies the network effect. A network effect is the effect of a network's value V is dependent on its size n (the number of its nodes)~\cite{zhang2015tencent}. Four laws have been proposed to provide more precise definitions and characterizations of network effect~\cite{zhang2015tencent}. They are Sarnoff's law~\cite{swann2002functional}, Odlyzko's law~\cite{briscoe2006metcalfe}, Metcalfe's law~\cite{metcalf1993metcalf} and Reed's law~\cite{reed1999sneaky}. Metcalfe himself~\cite{metcalfe2013metcalfe} used Facebook's data over the past 10 years to show a good fit for Metcalfe's law.  Zhang et al.~\cite{zhang2015tencent} expanded Metcalfe's results by utilizing the actual data of Tencent and Facebook and validated Metcalfe's Law. 

\section{The conclusion}

This article concluded three laws of technology: laws of technology inertia, technology change force, and technology action and reaction, and derived five corollaries: corollaries of measurement of technology change force, technology breakthrough, technology monopoly, technology openness, and technology business opportunities. These laws and corollaries provide a theory framework to explain or approximate technology rise or fall. I presented how to use this theoretical framework to analyze an emerging technology---RISC-V. Also, I elaborated on benchmarks' ro\-le in applying those laws.

\section{Acknowledgments}
I am very grateful to many persons' contributions, especially  Dr. Lei Wang for discussing three laws, contributing and discussing Table~\ref{tab1}, contributing Fig.~\ref{processor_ecosystem}, and proofreading the article, Mr. Shaopeng Dai for editing this article, drawing Fig.~\ref{FIG:1}-\ref{FIG:4} and compiling the references, Dr. Chunjie Luo and Dr. Wanling Gao for contributing and discussing Table~\ref{tab1} and proofreading this article.



\printcredits

\bibliographystyle{cas-model2-names}

\bibliography{cas-refs}


\bio{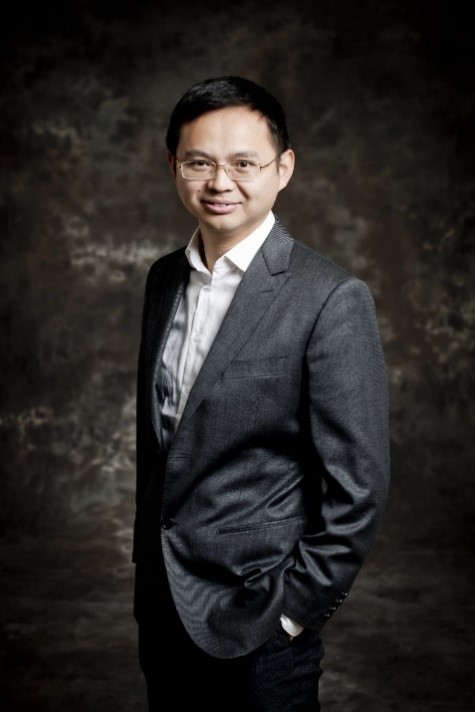}
Dr. Jianfeng Zhan is a Full Professor at Institute of Computing Technology (ICT), Chinese Academy of Sciences (CAS), and University of Chinese Academy of Sciences (UCAS), and the director of Research Center for Advanced Computer Systems, ICT, CAS.  He received his B.E. in Civil Engineering and MSc in Solid Mechanics from Southwest Jiaotong University in 1996 and 1999, and his Ph.D. in Computer Science from Institute of Software, CAS, and UCAS in 2002. His research areas span from Chips, Systems to Benchmarks. A common thread is benchmarking, designing, implementing, and optimizing a diversity of systems. He has made substantial and effective efforts to transfer his academic research into advanced technology to impact general-purpose production systems. Several technical innovations and research results, including 35 patents, from his team, have been adopted in benchmarks, operating systems, and cluster and cloud system software with direct contributions to advancing the parallel and distributed systems in China or even in the world. He has supervised over ninety graduate students, post-doctors, and engineers in the past two decades. 
Dr. Jianfeng Zhan founds and chairs BenchCouncil and serves as the Co-EIC of TBench with Prof. Tony Hey. He has served as IEEE TPDS Associate Editor since 2018. He received the second-class Chinese National Technology Promotion Prize in 2006, the Distinguished Achievement Award of the Chinese Academy of Sciences in 2005, and the IISWC Best paper award in 2013, respectively. \endbio
\end{document}